\documentclass{ceab}   

\usepackage{epsfig}     
\usepackage{hyperref} 
\usepackage{ceabbib}     
\usepackage[T1]{fontenc}
\usepackage{float} 

\begin{document}
\floatplacement{figure}{H} 	
\floatplacement{table}{H} 	

\title{Identification of coronal holes and filament channels in SDO/AIA 193\AA~images via geometrical classification methods}
\def\gore{M. Reiss et al.}
\def\tit{Coronal holes and filament channels in SDO/AIA images}

\author{M. REISS$^{1}$, M. TEMMER$^1$, T. ROTTER$^1$, \\S.~J. HOFMEISTER$^1$ and A.~M. VERONIG$^{1,2}$
\vspace{2mm}\\
\it $^1$Institute of Physics, NAWI Graz, University of Graz, Austria\\ 
\it $^2$Kanzelh\"ohe Observatory, University of Graz, Austria}

\maketitle

\begin{abstract}
In this study, we describe and evaluate shape measures for distinguishing between coronal holes and filament channels as observed in Extreme Ultraviolet (EUV) images of the Sun. For a set of well-observed coronal hole and filament channel regions extracted from SDO/AIA 193\AA~images we analyze their intrinsic morphology during the period 2011 to 2013, by using well known shape measures from the literature and newly developed geometrical classification methods. The results suggest an asymmetry in the morphology of filament channels giving support for the sheared arcade or weakly twisted flux rope model for filaments. We find that the proposed shape descriptors have the potential to reduce coronal hole classification errors and are eligible for screening techniques in order to improve the forecasting of solar wind high-speed streams from CH observations in solar EUV images. 
\end{abstract}


\section{Introduction}
 
It is well known that coronal holes (CH) play an important role in geomagnetic storm activity \citep{tsurutani06}. They coincide with rapidly expanding open magnetic fields and are the source regions of the high speed solar wind streams (HSSs). HSSs shape the solar wind distribution in interplanetary space and are the dominant contribution to space weather disturbances at times of quiet solar activity. Due to the lower temperature and density compared to the ambient coronal plasma, coronal holes are visible as dark areas in the solar corona as observed in EUV. In the past, coronal holes have mostly been visually identified visually and tracked by experienced observers. There have been a few attempts to automate the process for the identification and detection of coronal holes using automatic image recognition techniques \citep{delouille07, kirk09, krista09, detoma11, rotter12}. Those techniques are mainly based on differences in intensity compared to the ambient corona and little attention has been paid to the characteristic geometry of the features under study.

The automated method for identification and extraction of CH regions, used to forecast the solar wind speed at 1 AU developed by \cite{vrsnak07} and \cite{rotter12} is currently applied on SDO/AIA 193\AA~images \citep{lemen12}. To this aim we use a simple but powerful histogram-shape-based thresholding technique to obtain the fractional CH areas from SDO/AIA 193\AA~images. Fractional CH areas extracted at a 15$^\circ$ slice at the solar meridian reveal a distinct correlation with the solar wind speed measured $\sim$4 days later in-situ at 1~AU. A recent improvement of the method includes an automatically adapting relation between CH area and solar wind speed based on the three preceding Carrington rotations (Rotter et al., 2014; in preparation). The hourly updated results are published online at \url{http://swe.uni-graz.at/solarwind}.

\begin{figure}
\begin{center}
\epsfig{file=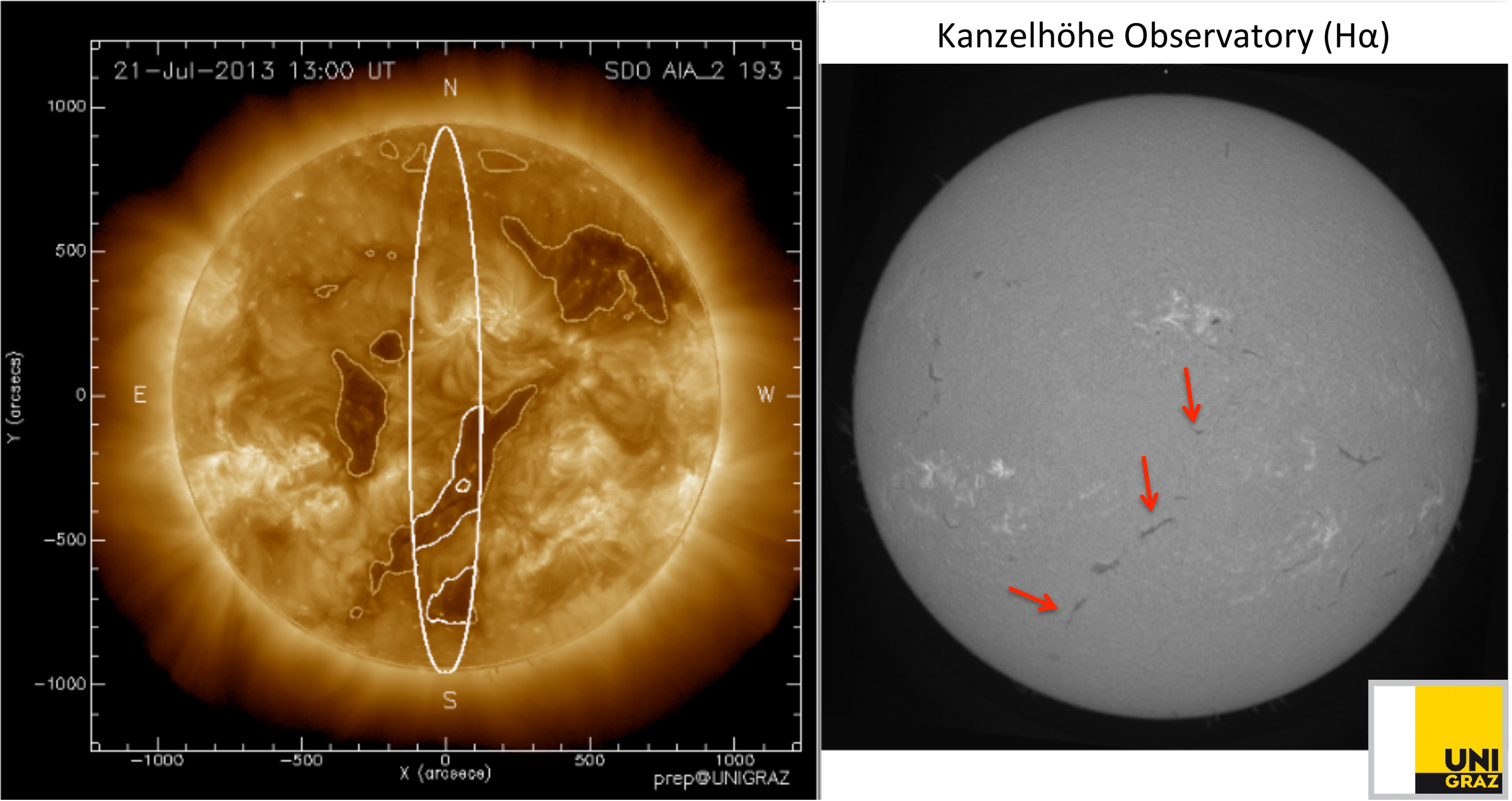,width=10cm}%
\end{center}
\vspace{-0.5cm}
\caption{\textit{Left:} Coronal holes identified in SDO/AIA 193\AA~observations from July 21, 2013 based on a histogram-shape-based thresholding technique \citep{rotter12}. \textit{Right:} H$\alpha$ data from the Kanzelh\"ohe Observatory of the University of Graz, Austria. The identified coronal hole area at the central meridian corresponds to an extended filament channel. }\label{f1}
\end{figure}

Due to the almost equally low intensity in solar EUV images, filament channels (FCs) are sometimes erroneously identified as CHs affecting the quality of the solar wind speed forecast. Hence, to improve the solar wind forecasting method we need to distinguish FCs from CHs in EUV images (cf.\, Fig.~\ref{f1}). We tackle this issue by the physical properties of FCs which is reflected in their geometric characteristics. Compared to CHs, FCs are very different in their magnetic field configuration as they are regions of closed magnetic field lines along a polarity inversion line. The FCs form prior to the actual visible filament due to the formation of a low lying flux rope of magnetic field. They are usually interpreted in terms of the sheared arcade or weakly twisted flux rope model, having a magnetic field which is dominated by the axial component. The dense/cold prominence material is located in the dip of the helical windings leading to the elongated dark structures as observed at EUV wavelengths \citep[for more details on the formation of filaments we refer to the review by][]{mackay10}. Thus, we act on the assumption that FCs are, compared to CHs, long narrow features that extend in a particular direction, which is also supported by H$\alpha$ observations (cf.\, Fig.~\ref{f1}).

Based on this geometric distinction, we aim to further improve the automatized CH identification method by implementing a geometrical parameter which separates FCs from CHs. The ultimate goal would be to determine a single characteristic value with respect to a distinct recognition of CHs and FCs in EUV images. We identify shape measures that can clearly be assigned to either of the features. In Sect.\,2 we analyze the eligibility of standard shape measures as found in the literature and two newly developed algorithms. In Sect.\,3 we apply the different measures to the observational data and present the results. In Sect.\,4 we discuss the findings and draw our conclusions in Sect.\,5.

\section{Methods}
Shape descriptors have become an important tool in pattern recognition and computer vision because they allow to characterize the shape of an object pixel configuration with a single scalar number. To reduce possible loss of shape information we focus on shape measures like compactness, roundness and elongation, which theoretically do not change with spatial resolution. 

\subsection{Established Methods}
The compactness, $c$, is a very popular shape measure and can be defined as $c = U^2/{4 \pi A}$, where $U$ is the perimeter, and $A$ is the object area, calculated by the sum over all pixels associated to the object. The perimeter is defined as the total length of the boundary of an object and can be calculated using different approaches, as e.g., the simple sum of all border cells or the sum of all distances between the centers of all boundary pixels. We detect the border cells with a widely applied border tracing algorithm \citep{imageprocessing98}. 

\vspace{0.25cm}

\begin{table}      
$$ 
\begin{array}{p{0.5\linewidth}l}
\hline
\noalign{\smallskip}
Shape Measure      &  Formula \\
\noalign{\smallskip}
\hline
\noalign{\smallskip}
Compactness           & c = U^2 / 4 \pi A \\
Roundness     & r = 1 / c    \\
Elongation & e = d_l / d_s    \\
\noalign{\smallskip}
\hline
\end{array}
$$ 
\caption[]{Established definitions for shape measures.}      
\end{table}

\vspace{0.25cm}
\noindent Intuitively one would expect that a less compact structure has a lower compactness. But exactly the opposite is the case. The most compact object in the continuous 2D space is a circular disk. In this case the factor $4 \pi$ ensures that $c=1$. This is the lowest possible value for the compactness. Thus, for less compact objects the value of $c$ increases. The reciprocal of the compactness is called the roundness of an object. Considering this definition a less rounded shape has a lower roundedness and vice versa. Both measures therefore demonstrate the relation of the shape to a perfect circle. The elongation is simply defined as the aspect ratio of the longest, $d_l$, and shortest distance $d_s$ of the bounding box of an object. The definitions of the standard shape measured used in this study are summarized in Table~I.

\subsection{Newly developed methods}
The shape measures described in Table~I reduce the object pixel configuration to a single scalar number, which hides some of the detailed shape information. In order to investigate the geometrical properties of CHs and FCs in detail, we propose two alternative shape measurements. 

\subsubsection{Symmetry Analysis}
\begin{figure}
\begin{center}
\epsfig{file=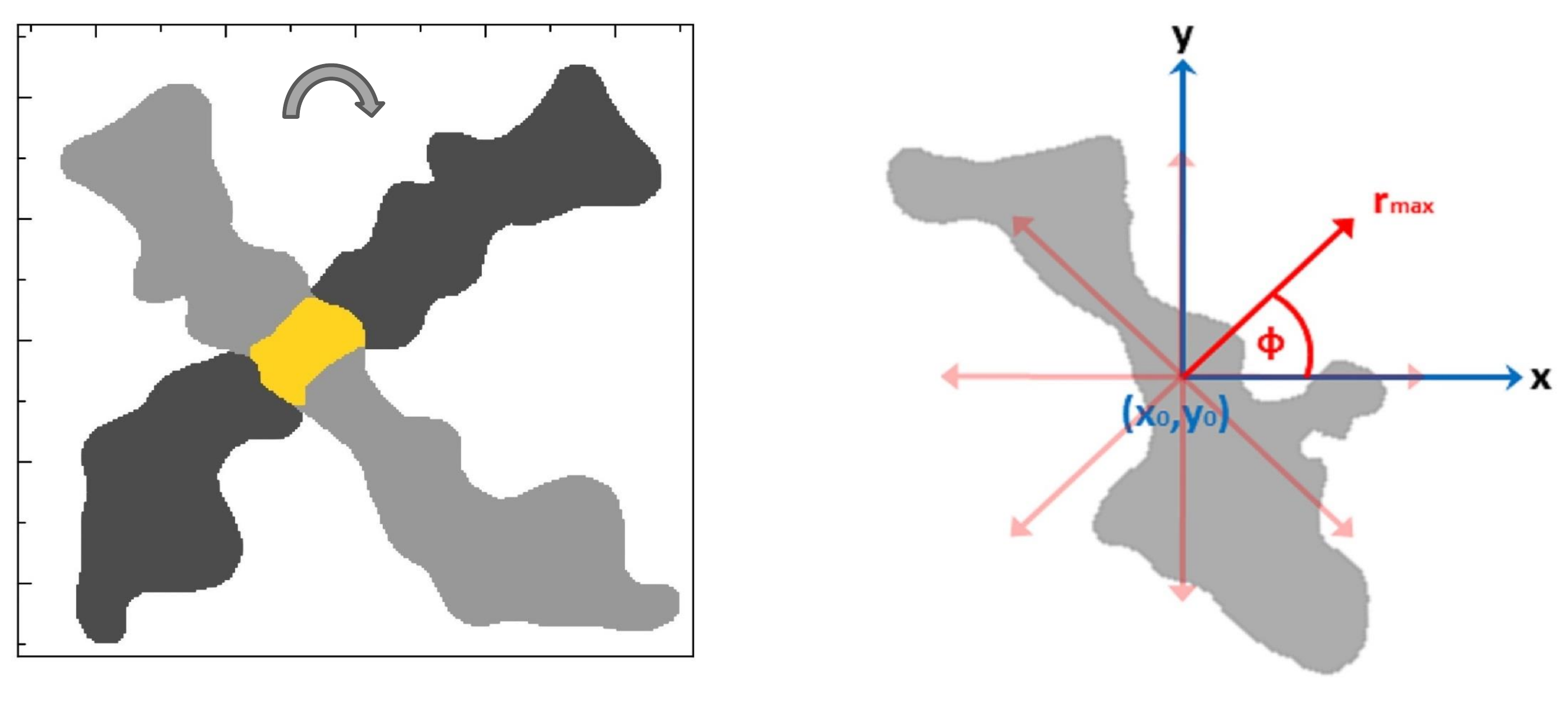,width=10cm}%

\end{center}
\vspace{-0.5cm}
\caption{Visualization of the developed classification methods. (left) Symmetry Analysis: Calculate the overlapping object pixels (yellow) in percentage after the application of a discrete transformation exemplary shown by using a rotation of 90$^\circ$. (right) Direction dependent analysis: Centre each object pixel $(x_0,y_0)$ and compute the average number of neighbours within angle $\phi$ and distance $r_{max}$.}\label{fig2}
\end{figure}

We investigate the geometrical symmetry properties by using the following technique. After the application of discrete geometrical transformations like rotation (in steps of 45$^\circ$), reflection and a composition of both, we sum over the overlapping object pixels and calculate the total overlap in percentage. A rotation of 360$^\circ$ would result in the maximum of overlap but does not provide new information as it represents exactly the same object. Therefore, as a characteristic shape measurement we use the average overlap in percentage of all transformations which provides different structures than
the original one (cf.\, left panel of Fig.~\ref{fig2}).

\subsubsection{Direction dependent analysis}
In a first step, we express the representative shape of an object in a functional form. As shown in the right panel of Fig.~\ref{fig2}, we center each pixel $(x_0,y_0)$ of an object and count the number of pixels within a given direction $\phi$ and distance $r_{max}$. Computing the average number of object pixels for each direction leads to the function $f(\phi)$. In order to compare CHs and FCs of different sizes we divide the average number of object pixels per direction by its maximal value as given by $f^*(\phi) = f(\phi) / f_{max}(\phi)$. In the next step we calculate the standard deviation of $f^*$, which provides a measure for the object shape. The division cancels the influence of the total number of object pixels and the standard deviation of the calculated function provides a shape specific value. 


\section{Results}

In order to obtain the most suitable geometrical method for our purpose, we analyze a set of 410 low intensity regions automatically extracted from SDO/AIA 193\AA~images over the time range 2011--2013 containing both structures, CHs and FCs. Only structures close to the central meridian were considered in order to reduce projection effects that would influence the true shape information. The structures are defined as discrete objects, i.e.\,pixels within the extracted boundary are set to $1$ and outside the boundary to $0$ (see also Fig.~\ref{fig2}). Based on the information obtained by visual inspection of H$\alpha$ filtergrams from Kanzelh\"ohe Observatory of the University of Graz, Austria \citep{poetzi13} we divide the set into 61 FCs and 349 CHs. This means that $\sim$15\% of automatically extracted areas are actually FCs. 

\begin{figure}
\begin{center}
\epsfig{file=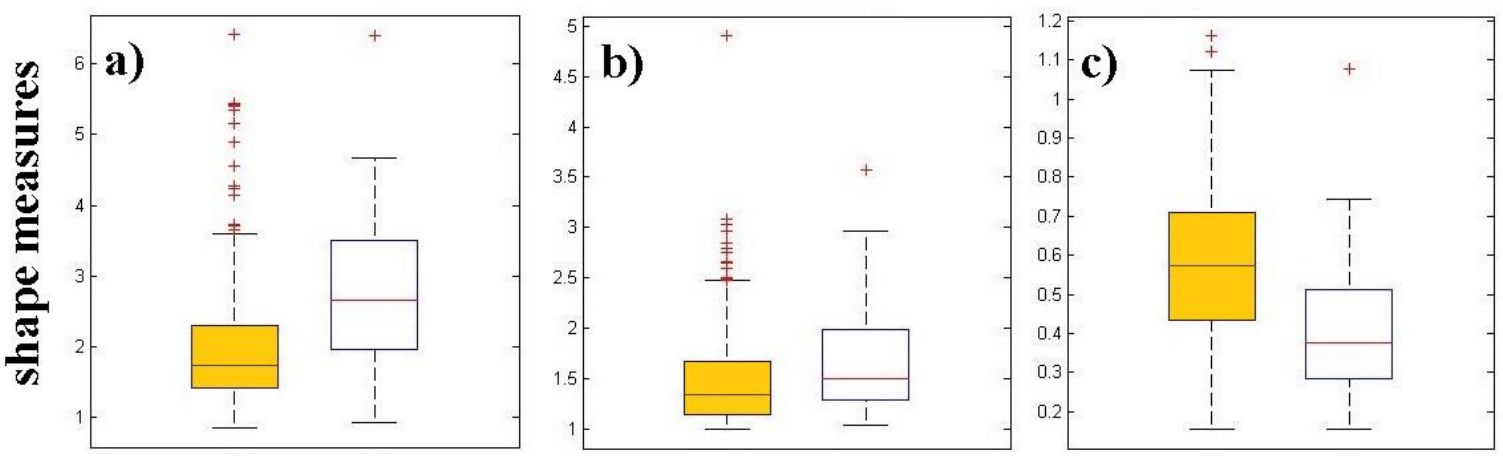,width=12cm}%
\end{center}
\vspace{-0.5cm}
\caption{Comparison between standard shape measures applied on structures of coronal holes (orange boxes) and filament channels (white boxes). On each box, the central mark is the median (50th percentile), the edges of the box are the 25th and 75th percentiles, the whiskers show the $\pm$2.7$\sigma$ range covering 99.3\% of data. Outliers are marked as red crosses. Methods: (a) elongation, (b) compactness, (c) roundness.}\label{fig3a}
\end{figure}

\begin{figure}
\begin{center}
\epsfig{file=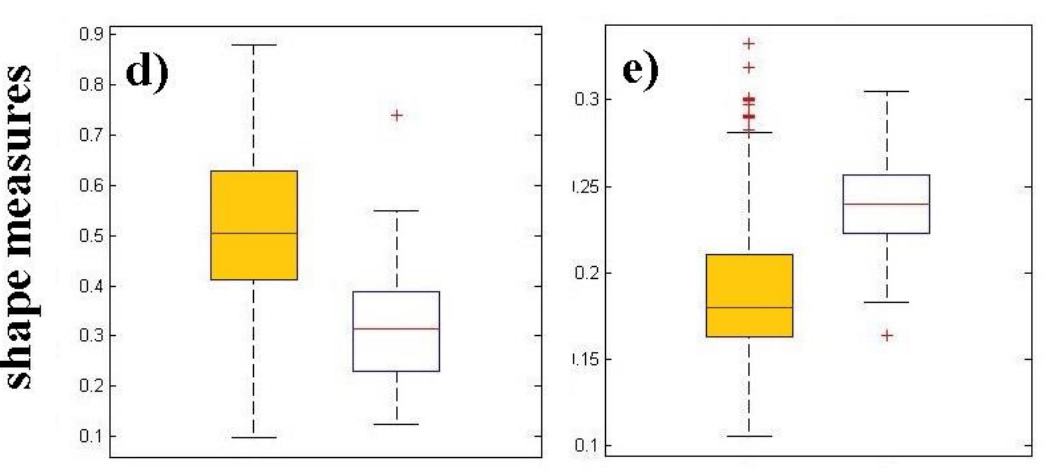,width=8cm}%
\end{center}
\vspace{-0.5cm}
\caption{Same as Fig.~\ref{fig3a} but for the proposed geometrical classification methods: (d) symmetry analysis, (e) direction dependent analysis.}\label{fig3b}
\end{figure}

In a next step we apply the different geometrical methods separately on the sets of CH and FC structures. Fig.~\ref{fig3a} and Fig.~\ref{fig3b} show the resulting shape measures when using the five different methods. The larger the differences between the boxes, the better the distinction between CHs and FCs. The plots in Fig.~\ref{fig3a} show that the well established methods, utilizing compactness, roundness and elongation, might not be successful for automatically distinguishing CHs and FCs as the differences between the measures are not very distinct. In fact the boxes show an overlap in the shape measures of $\sim$36\% for elongation, $\sim$35\% for compactness, and $\sim$26\% for roundness.

Fig.~\ref{fig3b} shows the results for the newly developed shape measures. It can be clearly seen that for these cases the distinction between CHs and FCs is much stronger and the derived overlaps are lower. We obtain an overlap of $\sim$20\% for the symmetry analysis and 17\% for the direction dependent analysis. Hence, the direction dependent analysis is obtained to be the most promising method. These results are not surprising since the study is based on the observational fact that FCs are in general long narrow features extending in a particular direction. We note that the results for the established methods also partially reflect the observed geometry of CHs and FCs, e.g., the elongation and compactness are larger for FCs compared to CHs whereas the roundness is found to be of the opposite trend.

\section{Discussion}
As shown in Fig.~\ref{fig3a} the elongation performs worst among the established methods. Obviously the ratio of the longest and the shortest distance of the minimal bounding box for an object is a too simple approach for distinguishing between two differently shaped structures. We obtain for the data set on which we apply the method an overlap of 36\% (fraction for which the shape measure gives the same result for FC or CH structure). 

With the concept of circularity and compactness we compare the actual shape of the object with that of a circle of the same size \citep{costa00}. Both methods have the disadvantage that the perimeter has enormous influence on the circularity and compactness because this shape measures are proportional to the square of the perimeter. This means for example the more details a circle shows the larger the value for the circularity because the perimeter increases quadratically whereas the area increases linearly \citep[for more details see][and references therein]{montero09}. 

Due to the strong dependence on the perimeter, the methods are not well suited for a comparison of irregular rough shapes of different sizes and we derive an overlap of 35\% and 26\%, respectively. Furthermore there is no consensus about the correct method for the calculation of the perimeter formed by a set of discrete object pixels. We tested two different methods for the calculation of the perimeter \citep{imageprocessing98} but found no improvement for the differentiation between CHs and FCs. 

A good differentiation was achieved with the proposed symmetry analysis technique (overlap of 20\%). This measure is equal to $1$ for perfect symmetric structures and approaches $0$ for asymmetric structures. The advantage of the method in comparison to circularity and compactness is its usefulness in a discrete space. Circularity and compactness are only theoretical concepts which do not hold in discrete images. As discussed a circular disk is the most compact or most rounded shape in the continuous 2D space. But the most compact shape in a discrete space is not a circular disk. In order to overcome these problems we propose to focus only on the symmetry properties of the discrete object pixel configurations. The proposed measure is neither dependent on the property of an object like the perimeter nor it is sensitive to this discretization problems. 

The best results were achieved with the direction dependent analysis of FCs and CHs (overlap of 17\%). The average number of neighboring pixels per direction is a characteristic shape function of the analyzed object pixel configuration. An alternative approach would be based on the center of mass of the object from which the number of neighboring pixels is calculated in each direction. However, this would strongly depend on the position of the center of mass relative to the border line, hence, would not be useful for the analysis of elongated irregular shapes.

A good compromise for this specific application is the use of the proposed overlapping scanning technique. The average number of neighboring pixel per direction has two contributors, the number of object pixels per direction and also the total number of object pixels within the structure. In order to focus on the specific shape we propose to divide this function by its maximum value and use the standard deviation of this calculated function as a shape descriptor.

\section{Conclusion}

The automatized CH extraction method which uses a histogram shape-based thresholding technique based on solar EUV images \citep{rotter12} sometimes erroneously identifies FCs due to their equally low intensities compared to CHs ($\sim$15\% of CHs are found to be FCs). For screening and cleaning the extracted low intensity areas, and with the goal to improve the forecasting of solar wind at 1~AU based on CH observations \citep{vrsnak07}, we apply geometrical methods. Considering the observational fact that FCs are largely asymmetric structures aligned along a preferred direction, we tested well known shape descriptors and also developed two new methods.

The methods are applied on a data set of CHs and FCs in SDO/AIA 193\AA~images recorded during the period 2011 to 2013, from which we evaluate their significance for distinguishing between CHs and FCs. All the discussed object measures are theoretically invariant if the object is moved or scaled. A clear trend is observable for all standard shape measures (elongation, compactness, roundness) but not distinct enough to be sufficient for this purpose. For the newly proposed methods we obtain that the direction dependent analysis and the symmetry analysis show the lowest overlap (20\% and 17\%, respectively), which makes it possible to distinguish FCs from CHs in an automated mode. Hence, those geometrical methods are found to be eligible for our purpose. Finally, we stress that the algorithms work adequately fast to be applied on real-time data for improving our automated solar wind forecast.

\section*{Acknowledgments} 
We gratefully acknowledge the NAWI Graz funding zur \textit{F\"orderung von JungforscherInnengruppen 2013--15}. T.\,R. is supported by the research grant \textit{Beihilfe f\"ur Zwecke der Wissenschaft}. The research leading to these results has received funding from the European Commission's Seventh Framework Programme (FP7/2007-2013) under the grant agreement no.~284461 [eHEROES]. 

\bibliographystyle{ceab}


\end{document}